\documentclass[showpacs,onecolumn,pra,notitlepage]{revtex4-1}
\usepackage[dvips]{graphicx}
\usepackage{amsmath,amssymb,amsthm,mathrsfs,amsfonts,dsfont}
\usepackage{subfigure, epsfig}
\usepackage{braket}
\usepackage{bm}
\usepackage{enumerate}
\usepackage{multirow}
\usepackage{color}
\usepackage{qcircuit}
\usepackage{lineno}
\usepackage{array}     
\usepackage{comment}
\usepackage{enumitem} 
\usepackage{epstopdf}
\usepackage{booktabs,colortbl} 
\usepackage{rotating}
\usepackage{tcolorbox}
\usepackage{hyperref}
\usepackage{ulem}
\usepackage{longtable}
\usepackage{setspace}

\graphicspath{{./figure/}}

\begin{document}

\title{Surpassing the rate-transmittance linear bound of quantum key distribution}

\date{\today}

\author{Xiao-Tian Fang$^{1,\,2,\,*}$}
\author{Pei Zeng$^{3,\,*}$}
\author{Hui Liu$^{1,\,2,\,*}$}
\author{Mi Zou$^{1,\,2}$}
\author{Weijie Wu$^{3}$}
\author{Yan-Lin Tang$^{4}$}
\author{Ying-Jie Sheng$^{4}$}
\author{Yao Xiang$^{4}$}
\author{Weijun Zhang$^{5}$}
\author{Hao Li$^{5}$}
\author{Zhen Wang$^{5}$}
\author{Lixing You$^{5}$}
\author{Ming-Jun Li$^{6}$}
\author{Hao Chen$^{6}$}
\author{Yu-Ao Chen$^{1,\,2}$}
\author{Qiang Zhang$^{1,\,2}$}
\author{Cheng-Zhi Peng$^{1,\,2,\,4}$}
\author{Xiongfeng Ma$^{3}$}
\author{Teng-Yun Chen$^{1,\,2}$}
\author{Jian-Wei Pan$^{1,\,2}$}

\affiliation{$^1$Hefei National Laboratory for Physical Sciences at Microscale and Department of Modern Physics, University of Science and Technology of China, Hefei, Anhui 230026, China}
\affiliation{$^2$CAS Center for Excellence in Quantum Information and Quantum Physics, University of Science and Technology of China, Hefei, Anhui 230026, China}
\affiliation{$^3$Center for Quantum Information, Institute for Interdisciplinary Information Sciences, Tsinghua University, Beijing 100084, China}
\affiliation{$^4$QuantumCTek Corporation Limited, Hefei, Anhui 230088, China}
\affiliation{$^5$State Key Laboratory of Functional Materials for Informatics, Shanghai Institute of Microsystem and Information Technology, Chinese Academy of Sciences, Shanghai 200050, China}
\affiliation{$^6$Corning Incorporated, Corning, New York 14831, USA}
\affiliation{$^*$These authors contributed equally to this work.}



\begin{abstract}
Quantum key distribution (QKD)\cite{Bennett1984Quantum,Ekert1991Quantum} offers a long-term solution to establish information-theoretically secure keys between two distant users. In practice, with a careful characterization of quantum sources and the decoy-state method\cite{Hwang2003Decoy,Lo2005Decoy,Wang2005Decoy}, measure-device-independent quantum key distribution (MDI-QKD)\cite{Lo2012Measurement} provides secure key distribution. While short-distance fibre-based QKD has already been available for real-life implementation\cite{tang2014field}, the bottleneck of practical QKD lies on the limited transmission distance. Due to photon losses in transmission, it was believed that the key generation rate is bounded by a linear function of the channel transmittance, $O(\eta)$, without a quantum repeater\cite{takeoka2014fundamental,Pirandola2017Fundamental}, which puts an upper bound on the maximal secure transmission distance\cite{Yin2016MDIQKD,Boaron2018secure}. Interestingly, a new phase-encoding MDI-QKD scheme, named twin-field QKD\cite{Lucamarini2018TF}, has been suggested to beat the linear bound, while another variant, named phase-matching quantum key distribution (PM-QKD), has been proven to have a quadratic key-rate improvement\cite{ma2018phase,Lin2018simple}, $O(\sqrt{\eta})$. In reality, however, the intrinsic optical mode mismatch of independent lasers, accompanied by phase fluctuation and drift, impedes the successful experimental implementation of the new schemes. Here, we solve this problem with the assistance of the laser injection technique and the phase post-compensation method. In the experiment, the key rate surpasses the linear key-rate bound via 302 km and 402 km commercial-fibre channels, achieving a key rate over 4 orders of magnitude higher than the existing results in literature\cite{Yin2016MDIQKD}. Furthermore, with a 502 km ultralow-loss fibre, our system yields a secret key rate of 0.118 bps. We expect this new type of QKD schemes to become a new standard for future QKD.
\end{abstract}

\maketitle


In conventional point-to-point QKD, such as the BB84 protocol\cite{Bennett1984Quantum}, the sender Alice encodes key information into quantum states and sends them to the receiver Bob for detection. Whereas in MDI-QKD, Alice's and Bob's positions are symmetric. They both send out encoded optical pulses to a measurement site owned by Charlie, who interferes the pulses and publicly announces the results to correlate Alice's and Bob's key information. The security of MDI-QKD does not depend on how Charlie realizes the measurement or announces the results. As a result, this scheme is immune to all attacks on the detection and hence owns a higher security level in practice.

In quantum communication, attenuated lasers are widely used as photon sources, which can be described by weak coherent states, $\ket{\alpha} = e^{-|\alpha|^2/2} \sum_{k=0}^{\infty} \frac{\alpha^k}{k!} \ket{k}$, superpositions of $k$-photon states $\ket{k}$. The parameter $\alpha = \sqrt{\mu} e^{i\phi}$ is a complex number, where $\mu=|\alpha|^2$ is the light intensity and $\phi$ is the phase. In the original MDI-QKD\cite{Lo2012Measurement}, the user encodes the key information into two weak coherent states on two orthogonal optical modes, such as polarization encoding. In the security analysis, only the information carried by the single-photon states can be used for the final key generation. The decoy-state method is widely employed to efficiently extract secret key information\cite{Hwang2003Decoy,Lo2005Decoy,Wang2005Decoy}.

In reality, quantum information carriers, photons, can be easily lost during transmission. Define the transmittance between Alice and Bob $\eta$ to be the probability of a photon being successfully transmitted through the channel and being detected. Hence, in the symmetric setting of MDI-QKD, the transmittance between Alice (Bob) and the measurement site is $\sqrt{\eta}$. Only the detections, caused by Alice's and Bob's single-photon states, can be used for secure key generation. Then such detection rate is given by $O(\eta)$, as a natural upper bound of the key rate. 

\begin{figure}[htbp]
\includegraphics[width=9cm]{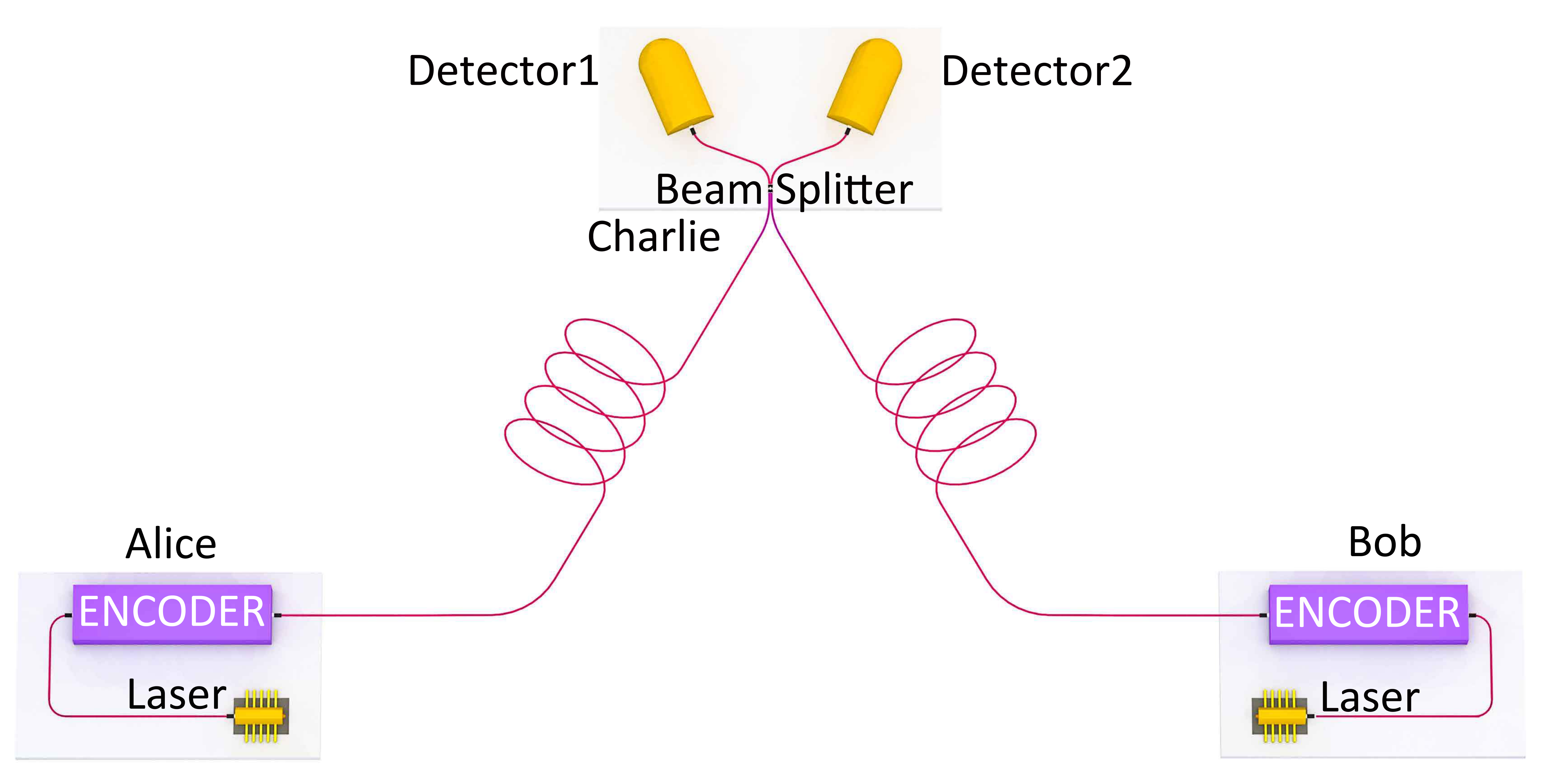}
\caption{\textbf{Schematic diagram of PM-QKD.} Encoder is a device to modulate the intensity $\mu$ and phase $\phi$ of coherent states. The beam splitter and the single photon detectors are used for interference detection. Divide the phases $\phi\in[0,2\pi)$ into $D$ slices, denoted by $\Delta_j=\left[\frac{\pi}{D}(2j-1), \frac{\pi}{D}(2j+1)\right)$ with the index $0\le j\le D-1$. In the experiment, we set $D=16$. In each round of key distribution, Alice encodes a random bit $\kappa_a$ into her coherent state $\ket{\sqrt{\mu_a} e^{i(\kappa_a\pi+\phi_a)}}$, after adding an extra discrete random phase $\phi_a = j_a 2\pi/D$ which is at the center of $j_a$th phase slice $\Delta_{j_a}$. Similarly, Bob encodes $\kappa_b$, $\mu_b$, and $\phi_b$ on his pulse, $\ket{\sqrt{\mu_b} e^{i(\kappa_b\pi+\phi_b)}}$. Alice and Bob then send their pulses to Charlie, who is supposed to interfere these quantum states to measure phase differences. After Charlie announces the detection results, Alice and Bob publicly announce the slice indexes $j_a, j_b$ of the random phases. They post-select the key bits $\kappa_a, \kappa_b$ as the raw key, according to Charlie's detection results and the sifting scheme depending on $j_a, j_b$, with the phase post-compensation technique.} \label{fig:PMQKD}
\end{figure}

To achieve a better rate-transmittance performance, a new phase-encoding MDI-QKD scheme, named PM-QKD, has been proposed, as shown in Fig.~\ref{fig:PMQKD}. Alice (Bob) encodes the key information into the phase of a coherent state on a single optical mode. In this case, Charlie treats Alice and Bob's two optical modes as one quantum system and detects the relative phase between them. To do so, Charlie only needs one photon in the joint quantum system. Therefore, the detection rate is $O(\sqrt\eta)$. Strict security analysis shows that PM-QKD enjoys a quadratic improvement on the rate-transmittance performance over the original MDI-QKD\cite{ma2018phase}.

Despite the promising qualities of PM-QKD on both security and performance, the experimental implementation is very challenging. In PM-QKD, the interference results at the measurement result should reflect the difference between Alice's and Bob's encoded phases. In practice, an essential requirement is to match the phases of coherent states generated by two remote and independent lasers. The coherent states of Alice $\ket{\sqrt{\mu_a} e^{i(\kappa_a\pi+\phi_a)}}$ and Bob $\ket{\sqrt{\mu_b} e^{i(\kappa_b\pi+\phi_b)}}$ could have different phase references due to phase drift and fluctuation. Define the reference deviation $\phi_\delta$ to be the phase difference when both Alice and Bob set $\kappa_{a(b)} = 0$ and $\phi_{a(b)}=0$. There are three main factors determining the value of $\phi_\delta$, fluctuations of the laser initial phases, fibre lengths, and laser frequencies. Take the $1550$ nm telecom light through a $200$ km fibre for example, either a tiny change of transmission time, say by $10^{-15}$ s corresponding to a 200 nm optical length, or a small deviation of the angular frequency, say by 1 kHz, will cause a significant phase drift. Note that, there are several recent experiments which make efforts to deal with these challenges in order to demonstrate the advantages of the new type of MDI-QKD schemes\cite{minder2019experimental,liu2019experimental,wang2019beating,zhong2019proof}.

\begin{figure*}[htbp]
\includegraphics[width=14cm]{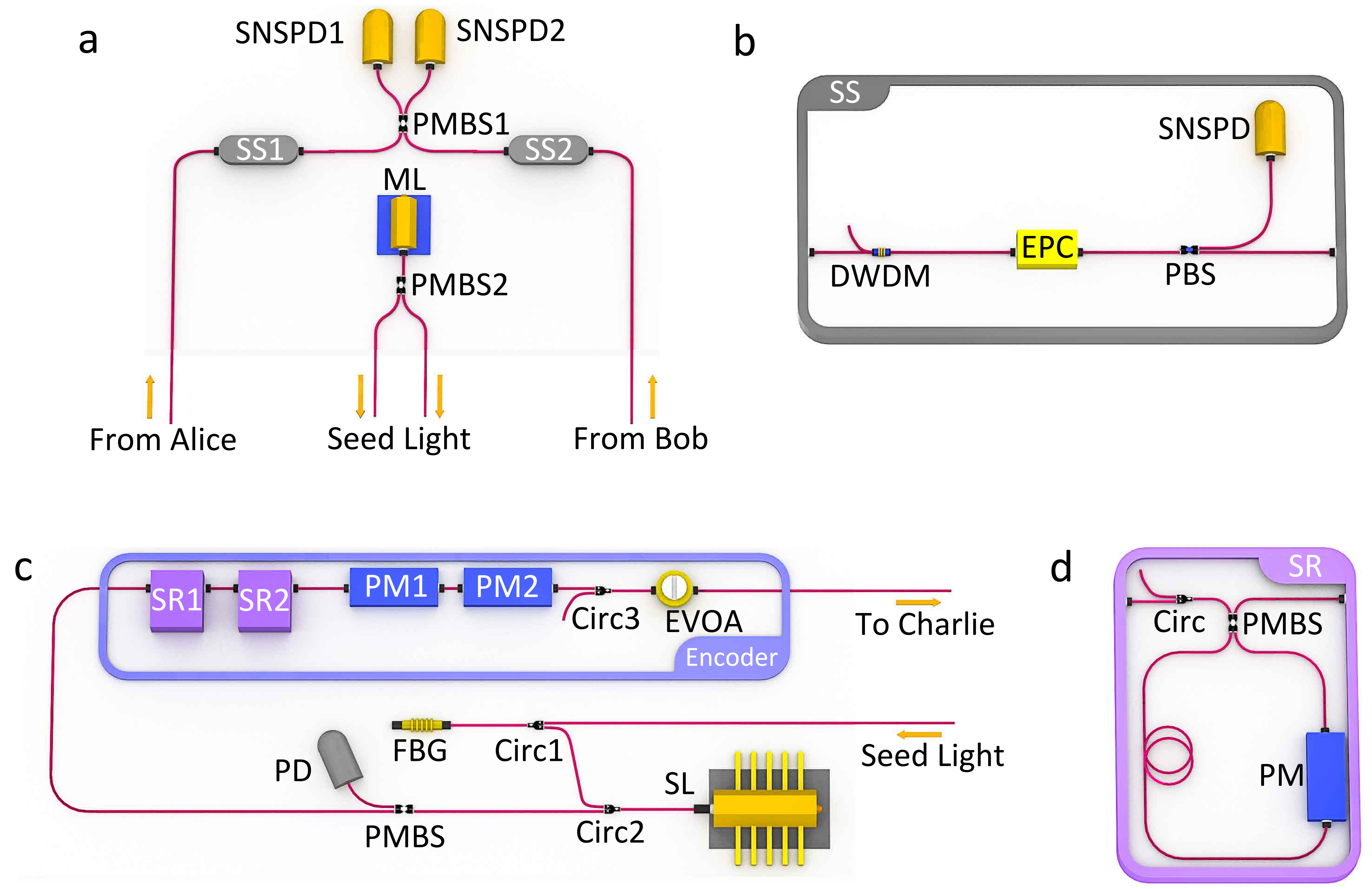}
\caption{ \textbf{Experimental setup.} \textbf{a.} The continuous-wave master laser (ML), used as the phase and wavelength reference, are split by a polarization-maintaining beam splitter (PMBS2) and sent to Alice and Bob to lock their distributed feedback (DFB) lasers, which act as slave lasers (SLs). Two stabilization systems (SS1 and SS2) are placed to enhance the interference stability. Alice and Bob’s pulses are interfered at PMBS1 and then detected by two superconducting nanowire single photon detectors (SNSPD1 and SNSPD2).  \textbf{b.} Stabilization system. The dense wavelength division multiplexer (DWDM) filters out optical noises which disturb the detection results. A polarization beam splitter (PBS), an SNSPD, and an electric polarization controller (EPC) are used to ensure that the polarization of the two pulses from Alice and Bob are indistinguishable. \textbf{c.} Quantum source. Alice (same as Bob) injects the seed light from the ML, which is filtered by the fibre Bragg grating (FBG), into her local DFB laser as the slave laser (SL). The SL generates pulses, which are split by the PMBS. One of the output pulses goes to the encoder and the other is monitored by a photoelectric diode (PD). The encoder is composed of two Sagnac rings (SR1 and SR2) for modulating intensities and two phase modulators (PM1 and PM2) for encoding phases. A circulator (Circ3) is placed to isolate the source from the channel. The electrical variable optical attenuator (EVOA) reduces the pulse intensity down to the single-photon level. \textbf{d.} Sagnac ring. An SR includes a Circ, a PMBS, a PM, and optical fibre. The length difference between the optical fibre connected the output and input of the phase modulators are delicately designed to meet two times of the repetition frequency of the pulses.} \label{fig:setup}
\end{figure*}

In this work, we implement PM-QKD with the setup shown in Fig.~\ref{fig:setup}. In order to suppress the fluctuations of the laser initial phases and frequencies, we employ the laser injection technique\cite{Lipka2017optical}. The setups on Alice's and Bob's sides are exactly the same. Below, we take Alice's side for example. The master laser in between Alice and Bob, with 3~kHz line-width and 1550.12~nm centre wavelength, emits a seed light which goes through a long fibre to lock Alice's distributed feedback laser.
Alice's laser generates optical pulses with a clock rate of 312.5~MHz. Two Sagnac rings are employed to modulate pulses into four different intensities. The pulses with the largest intensity are used as reference pulses for phase estimation, while the other three pulses are used as the signal state, weak decoy state, and vacuum state to implement the decoy-state method. The extinction ratio between the signal state and the vacuum state is about 20 dB. Two phase modulators are employed to modulate 16 different phases. Details of the laser injection technique and the phase estimation are presented in Appendix \ref{Sc:PhaseEst} and \ref{Sc:LaserInj}.

The pulses from Alice and Bob are transmitted through long optical fibres and interfered at the measurement site. The interference results are detected by two superconducting nanowire single photon detectors. The dark count is about $10$ counts per second and the detector efficiency is about $40\%$. The total detection efficiency reduces to about $23\%$ owing to 1.2 dB insertion loss and $25\%$ non-overlapping between signal and detection windows. Two stabilization systems are inserted before interference to filter out the noise caused by the nonlinear effect of the fibre and stabilizing the incident pulses. Details of the implementation are presented in Appendix \ref{Sc:ExpDetail}.

Due to fibre fluctuation, there exists a slow phase drift between Alice and Bob. In the case when the phase fluctuates slowly, pulses nearby share similar values of the reference deviation $\phi_\delta$. Inspired by this observation, we employ a simple phase post-compensation technique\cite{Ma2012Alternative}. During the experiment, Alice and Bob send reference pulses and quantum pulses periodically. The reference pulse is typically more than one order of magnitude stronger than the quantum pulse. They use reference pulses to estimate which slice $j_\delta$ the reference deviation $\phi_\delta$ lies in, according to interference results. They use quantum pulses to perform the PM-QKD experiment. After obtaining measurement results, Alice and Bob publicly announce the random phase slices $j_a, j_b$ of the signal pulses. They calculate $j_s=j_a-j_b+j_\delta$ for raw key sifting, where $j_\delta$ works as the post-compensation shift. Clearly, if $j_\delta$ accurately reflects the real-time reference deviation of the system, perfect interference will happen when $j_s=0$ or $j_s=8$. Note that the estimated phase slice indexes $j_\delta$ are only used in the post-processing, which frees us from active phase locking during the state transmission. Furthermore, compared with active locking, where the phase can only be locked well when $\phi_\delta$ keeps stable during the whole process of phase estimation and feedback, the phase post-compensation method can tolerate faster fluctuation, as long as $\phi_\delta$ does not change much in the time between the reference and quantum pulses.

In the security analysis of PM-QKD\cite{ma2018phase}, due to the random phases Alice and Bob modulated on their coherent states, the joint state can be decomposed into odd and even total photon number components, $\rho^{odd}_{AB}$ and $\rho^{even}_{AB}$. Denote the proportions of detection caused by $\rho^{odd}_{AB}$ and $\rho^{even}_{AB}$ as $q^{odd}$ and $q^{even}$, respectively, and obviously, $q^{odd} + q^{even} = 1$. 
In fact, the information leakage in PM-QKD is shown to be independent of channel disturbance. As a result, the privacy is only related to $q^{even}$, irrelevant to the bit error rate. The final key length is given by
\begin{equation} \label{eq:key}
K = M_\mu [ 1 - H(q^{even}_\mu) ] - l_{cor},
\end{equation}
where $H(x)=-x\log_2 x-(1-x)\log_2 (1-x)$ is the binary entropy function. Here, we consider the case where Alice and Bob use the same intensities of coherent states. The subscript $\mu$ represents the signal states with $\mu_a = \mu_b = \mu/2$. The raw key length $M_\mu$ is the number of detection events caused by signal states when Alice and Bob match their phases, $j_s=0$ or $8$. The even photon component ratio $q^{even}_\mu$ can be efficiently estimated by decoy state methods\cite{Hwang2003Decoy,Lo2005Decoy,Wang2005Decoy}. The error correction cost $l_{cor}$ can be usually estimated by a function of bit error rate $E_\mu$, $l_{cor} = f M_\mu H(E_\mu)$, where $f$ is the error correction efficiency depending on $E^{j_s}_\mu$. The key rate is defined by $R = K/N$, where $N$ is the number of QKD rounds. Details of the decoy state method and the security analysis by considering the finite data size effects are presented in Appendix \ref{Sc:Finite}.

In order to further improve the key rate, one can take advantage of the data with mismatched phases. Note that the phase-mismatched signals of $j_s = 1, 9$ can be regarded as the ones with a fixed misalignment $\phi_\delta = 2\pi/D$, which results in a larger bit error rate compared with the phase-matched signals of $j_s = 0, 8$. The raw keys with different $j_s$ have the same $q^{even}$ in Eq.~\eqref{eq:key} and hence the same privacy. More explicitly, Alice and Bob can categorize the data from signal states into $D/2$ groups, where the data of $j_s = 0, 8$ are in the $0$-th group, the data of $j_s = 1, 9$ are in the $1$-st group, etc. Alice and Bob can correct errors in each data group separately and perform privacy amplification altogether. Of course, if the error rate in a group is too large, they can simply discard that group of data.

\begin{figure*}[htbp]
\includegraphics[width=16cm]{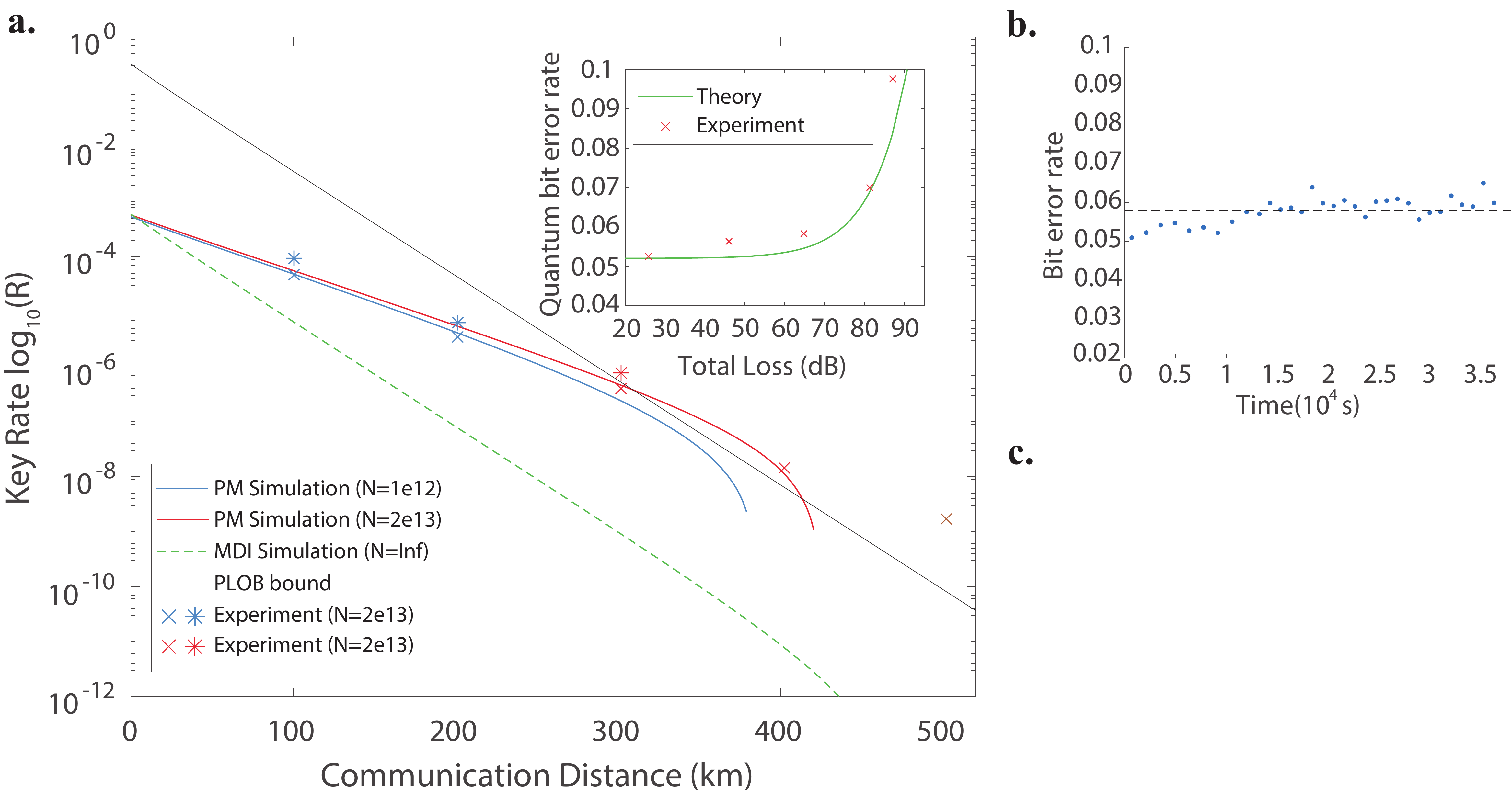}
\caption{\textbf{Experimental parameters and results.} \textbf{a.} Main panel, the experimental rate-distance performance of PM-QKD, comparing with the theoretical expectation and the linear key rate bound \cite{Pirandola2017Fundamental}. The data points marked by cross and star are, respectively, the key rate without and with the usage of phase-mismatched signals. Inset, error rate against total loss. The solid line shows the bit error rate in the theoretical model, and the crosses show the experiment data. \textbf{b.} The bit error rate with respect to the experiment time in the $302$ km experiment. Each data point represents a number of $3.31\times10^{5}$ effective clicks which are collected in $21.91$ minutes on average. The blue dots and red dots show the bit error rate for data with phase difference $j_d = 0$ and $1$, respectively. \textbf{c.} Parameters used in the experiment and theoretical simulation. Note that, the listed value of the dark count rate $p_d$, the detection efficiency $\eta_d$, misalignment error $e_d$, and the fibre loss of PM-QKD are used for numerical simulation. The corresponding experimental values depend on the specific environment, which are listed in Appendix \ref{Sc:ExpData}.} \label{fig:keyplot}
\end{figure*}

We perform the experiment via $101$, $201$, $302$, $402$ km standard optical fibres and a $502$ km ultralow-loss optical fibre. The experiment parameters and results are presented in Fig.~\ref{fig:keyplot}, from which one can see that the key rate-transmittance relation follows $R=O(\sqrt\eta)$ as a contrast with the linear rate-transmittance bound. Specifically, the experimental results beat the linear bound at the distances of $302$ and $402$ km. 
Take the 302 km fibre case as an example, with the same channel transmission and detection efficiency, the linear key-rate bound is given by $R_{up} = 5.44\times 10^{-7}$. Our experiment yields a key rate of $R = 6.74\times 10^{-7}$, with a failure probability of $\epsilon = 1.68\times 10^{-10}$, when all the mismatched data is used. The key rate is $24.0\%$ higher than the bound. In the case of 302 km, the data with mismatched phases has a significant contribution to the overall key rate, which is $72.6\%$ larger than the value with only the phase-matched group considered. Notably, our achieved key rate is three orders of magnitude higher than the asymptotic key rate of the original MDI-QKD scheme\cite{Lo2012Measurement}.

Meanwhile, we obtain a positive key rate at $502$ km experiment with an ultralow-loss optical fibre, beating the current record of 421 km fibre communication distance of QKD\cite{Boaron2018secure}. The channel loss of $502$ km experiment is 81.7 dB and the total loss is 87.1 dB. This new loss-tolerance record is comparable with the high-orbital satellite link loss in free space.

Our results show that PM-QKD system is stable and economical, which facilitates the promotion of practical QKD. In the future, we expect to use the phase post-compensation technique to keep the system robust by increasing the system repetition frequency and enhancing the detectors' performance. Also, we expect that the design of PM-QKD experiment will be helpful for the construction of quantum repeater\cite{EntSwap1993,Briegel1998Repeater}, as well as extending the reach of the quantum internet.


\section*{Funding Information}
This work has been supported by the National Key R\&D Program of China (Grant No.~2017YFA0303903 and No.~2017YFA0304000), the Chinese Academy of Science, the National Fundamental Research Program, the National Natural Science Foundation of China (Grant No.~11875173, No.~61875182 and No.~11674193), Anhui Initiative in Quantum Information Technologies, and Fundamental Research Funds for the Central Universities (WK2340000083).

\section*{Acknowledgments}
We acknowledge H.~Zhou for the insightful discussions.

\section*{Author Contributions}
All authors contributed extensively to the work presented in this paper.


\section*{Author Information}
Reprints and permissions information is available at www.nature.com/reprints. The authors declare no competing financial interests. Readers are welcome to comment on the online version of this article at www.nature.com/nature.

\nolinenumbers

\begin{appendix}

\section{Phase drift estimation} \label{Sc:PhaseEst}
\begin{figure*}[htbp]
\includegraphics[width=12cm]{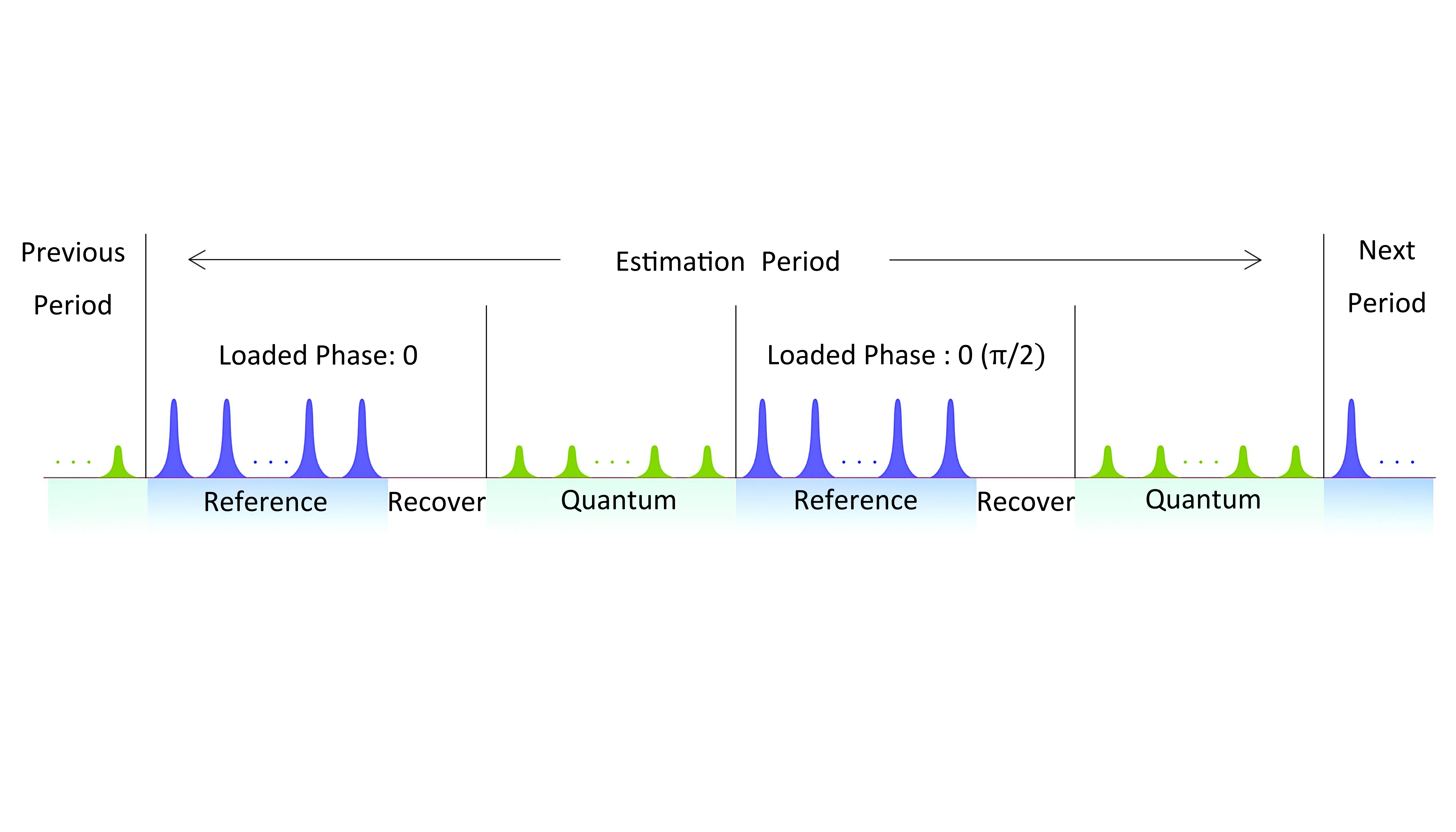}
\caption{\textbf{Phase estimation technique.} The pulses with a repetition rate of $312.5$ MHz are grouped into trains, each containing $625$ pulses. The period of pulse trains is about $2$ $\mu$s. In each train, there are two reference pulse regions and two quantum pulse regions. Alice loads the $0$ and $\pi/2$ phase on her pulses in the former and latter reference pulse region, respectively. A recovery region without pulses is at the end of every reference pulse region, where the detector systems recover to the optimal working condition.} \label{fig:postcomp}
\end{figure*}

Rather than using extra devices to lock the phase, here we apply the phase estimation method to estimate the drifted phase, as shown in Fig.~\ref{fig:postcomp}. Here, Alice and Bob need not obtain the exact value of the real-time phase deviation $\phi_\delta$, but only an estimation of the slice number $j_\delta$ for post-compensation. Moreover, the estimation of $j_\delta$ does not need to be announced in a real-time manner. Instead, it can be announced during the sifting process, as a post-selection shift factor. This makes our protocol practical without active feedback.

In the reference pulse region, Alice and Bob send reference pulses to Charlie, who interferes them and announces the results. They use the interference results to estimate the phase slice difference between two reference pulses, $\phi_\delta$. The right detector click ratio $P_r$ after interference is
\begin{equation}
\frac{n_r}{n_r + n_l} \approx P_r = \frac{1+\cos\phi_\delta}{2},
\end{equation}
where $n_r$ and $n_l$ are the counts of the right and left detector clicks, respectively. With this ratio, one can classify the phase fluctuation $\phi_\delta$ to the phase slices $\Delta_{j_\delta}$ marked by $j_\delta$ according to the detection ratio.

\begin{figure}[htbp]
\includegraphics[width=6cm]{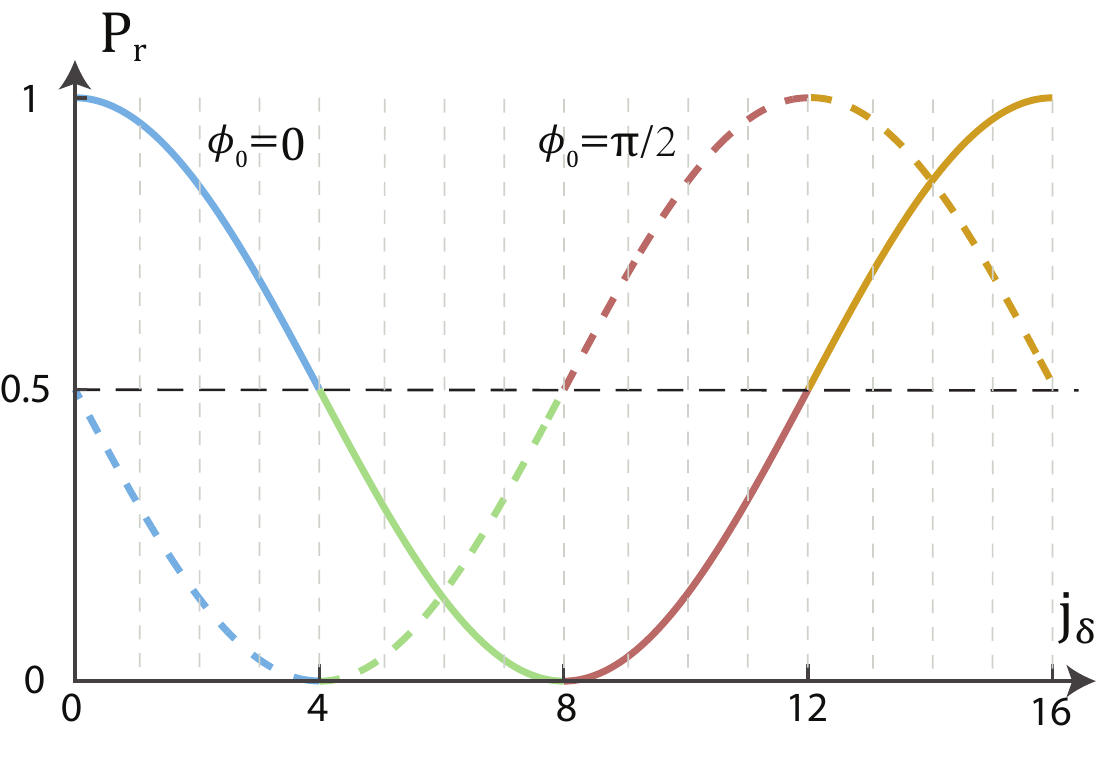}
\caption{\textbf{Phase estimation region.} The blue, green, red and orange region denotes four phase quadrants ranging from $[0,\pi/2)$, $[\pi/2,\pi)$, $[\pi,3\pi/2)$, and $[3\pi/2,2\pi)$. The solid line shows the theoretic detection probability $P_r$ with the initial phase difference $\phi_0$ of the reference pulse is set to $0$, while the dashed line shows $P_r$ when $\phi_0=\pi/2$. From these two estimated $P_r$ values, one can infer the $j_\delta$ value with no degeneracy.} \label{fig:phasequad}
\end{figure}

Note that, since the phase deviation $\phi_\delta$ and $(2\pi-\phi_\delta)$ yield the same $P_r$, Alice and Bob cannot discriminate these two cases from the ratio $P_r$. To solve this problem, Alice loads a $\phi_0 = \pi/2$ phase on the pulses in the latter reference pulse region, making the phase difference to be $\phi_\delta+\pi/2$, and hence $P_r = \frac{1-\sin\phi_\delta}{2}$. In that case, one can distinguish the phase slice $j_\delta$ from $(16-j_\delta)$. With the interference results $P_r$ from the case $\phi_0 = 0$ and $\phi_0 = \pi/2$, Alice and Bob can estimate $j_\delta$ accurately, as shown in Fig.~\ref{fig:phasequad}.

In order to yield an accurate estimation of $j_\delta$, sufficient detection counts of the reference pulses are required. According to the transmittance and phase drift velocity, one should properly set the intensity and time duration of the reference pulse and the system repetition frequency.

\section{Laser injection technique} \label{Sc:LaserInj}

The fluctuation of reference deviation $\phi_\delta$ is mainly caused by three factors: the initial phase fluctuation of the lasers, the optical length fluctuation, and the fluctuation of the laser frequencies. In order to minimize the fluctuation caused by the first and the third factors, we apply the laser injection technique.


\begin{figure}[htbp]
\includegraphics[width=8cm]{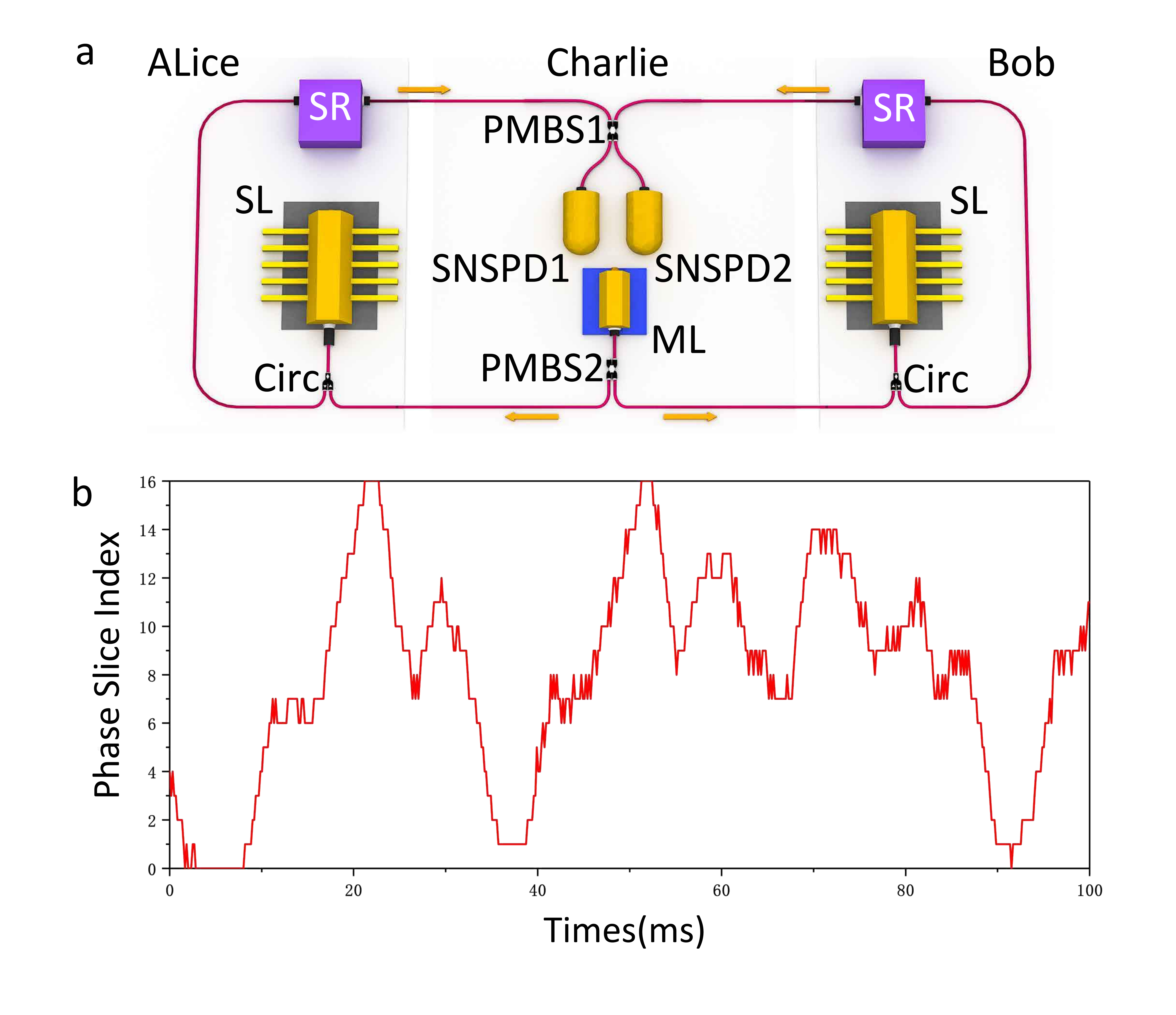}
\caption{\textbf{Laser injection technique and test.} \textbf{a.} Schematic implementation. \textbf{b.} Local drift test of the phase difference $\phi_\delta$ with two local slave lasers. A statistic analysis on the detection data shows that the phase difference between the two locked slave lasers drifts at $0.62$ rad/ms.}
\label{fig:laserinj}
\end{figure}


As shown in Fig.~\ref{fig:laserinj}a, a narrow line-width continuous wave laser at Charlie side works as the master laser, while DFB laser at Alice's (same as Bob's) side works as the slave laser. The seed light generated by the master laser is divided into two parts and sent to the two slave lasers through long fibres, which induces stimulated emission. In this case, the wavelength and phase of the light generated by the slave laser are the same as the seed light, which results in slower fluctuation of laser initial phases. For a local experiment test, interference results of pulses generated by slave lasers shows that the phase difference of two locked slave lasers fluctuates in a relatively low speed, which is presented in Fig.~\ref{fig:laserinj}b. Here, the residual phase noise and fluctuation mainly come from the spontaneous emission in the slave lasers. 
Though the phase will fluctuate faster when the fibre gets longer, we can still get an effecive phase estimation in this case.

In the $101$, $201$, $302$ km experiments, the master laser is put at the middle point between Alice and Bob's laser. The fibres used to transmit seed light and the signal light are different but with the same length. Due to the limited power of the master laser and low transmittance value, in the $402$ km and $502$ km experiments, we place the master laser and the slave lasers locally.

\section{PM-QKD protocol} \label{Sc:protocol}
In Box.~\ref{box:PMQKD}, we explicitly present the phase-matching quantum key distribution (PM-QKD) scheme implemented in the experiment. We divide the phase space $0\sim2\pi$ into $D$ slices and denote the $j$-th slice as $\Delta_j = [ \frac{\pi}{D}(2j-1), \frac{\pi}{D}(2j+1) )$. The central phase of $j$-th slice is $\phi_j = \frac{2\pi}{D}j$.

\begin{tcolorbox}[float, colback=orange!0!white,colframe=red!80!green,coltitle=white,title=Box 1 $|$ PM-QKD protocol, arc = 0mm,left=0mm,right=0mm, fonttitle=\sffamily\bfseries\large] \label{box:PMQKD}
\begin{itemize}[label={}]
\item
\textbf{State preparation}:
Alice randomly generates a key bit $\kappa_a$, and picks a random phase $\phi_{j_a}$ from the set $\{j\dfrac{2\pi}{D}\}_{j=0}^{D-1}$, and the intensity $\mu_{i_a}$ from the set $\{0,\nu/2,\mu/2\}$. She then prepares the coherent state $\ket{\sqrt{\mu_{i_a}}e^{i(\phi_{j_a}+\pi\kappa_a)}}_{A}$. Similarly, Bob generates $\kappa_b$, $\phi_{j_b}$, $\mu_{i_b}$ and then prepares $\ket{\sqrt{\mu_{i_b}}e^{i(\phi_{j_b}+\pi\kappa_b)}}_{B}$.

\item
\textbf{Measurement}: Alice and Bob send their optical pulses, $A$ and $B$, to an untrusted party, Eve, who is expected to perform an interference measurement and record the detector ($L$ or $R$) that clicks.

\item
\textbf{Announcement}: Eve announces her detection results. Then, Alice and Bob announce the random phases and intensities $j_a, \mu_{i_a}$ and $j_b, \mu_{i_b}$, respectively.

\item
\textbf{Phase estimation}: From time to time, Alice and Bob generate strong coherent pulses with $\phi_{j_a} = \phi_{j_b} = 0$ and send to Eve. Eve is supposed to interfere these pulses and estimate the phase difference $\phi_\delta$ between these two pulses. Eve announces the phase slice number $j_\delta$ of $\phi_\delta$.

Alice and Bob repeat the above steps for $N$ times. After that, they perform the following data post-processing procedures.

\item
\textbf{Sifting}: When Eve announces a successful detection, (a click from exactly one of the detectors $L$ or $R$), Alice and Bob keep $\kappa_a$ and $\kappa_b$. Bob flips his key bit $\kappa_b$ if Eve's announcement was an $R$ click. Then, Alice and Bob group the signals by $j_s = (j_a - j_b + j_\delta) \textit{ mod } D$. If $j_s \in [\frac{D}{4}, \frac{3D}{4})$, Bob flips his key bit $\kappa_b$. After that, Alice and Bob merge the data with $j_s$ and $j_s +\frac{D}{2}$, with $j_s = 0,1,..., \frac{D}{2}-1$.

\item
\textbf{Parameter estimation}: For all the raw data that they have retained, Alice and Bob record the detect number $M_{i_a, i_b}^{(j_s)}$ of different intensity combinations $\{\mu_{i_a}, \mu_{i_b}\}$ and phase group $j_s$. They then estimate the phase error number among all the clicked signal rounds $n^{X}_{\mu}$ in the phase group $j_s$ using the methods in Section \ref{Sc:decoy}.

\item
\textbf{Key distillation}: For the signals with $\mu_{i_a} = \mu_{i_b} = \frac{\mu}{2}$, Alice and Bob group them by the phase difference $j_s$. They then perform error correction and error verification on the raw key data of each group $j_s$, respectively. They then perform privacy amplification on the sifted key bits to generate private key.

\end{itemize}

\end{tcolorbox}

Suppose the phase references of Alice and Bob are different $\phi^{(0)}_a, \phi^{(0)}_b$. Moreover, we denote the phase variation in the fibre as $\phi_f$. Before Eve's interference, the phase difference between Alice and Bob's pulses is
\begin{equation}
(\phi_a + \phi_a^{(0)}) - (\phi_b + \phi_b^{(0)}) + \phi_f = \phi_a - \phi_b + (\phi_a^{(0)} - \phi_b^{(0)} + \phi_f) = \phi_a - \phi_b + \phi_\delta \approx (j_a - j_b + j_\delta) \frac{2\pi}{D},
\end{equation}
and hence the phase difference belongs to the $(j_a - j_b + j_\delta)$-th slice. Suppose $\phi_\delta$ fluctuates slowly with respect to time. By the interference of coherent pulses, Alice and Bob can estimate $\phi_\delta$ accurately. In this case, for the data with $j_s=0$, the interference should be nearly perfect, with small quantum bit error rate. Note that, the sifting strategy does not affect the security of PM-QKD. It only affects the error correction efficiency.

\section{Experiment detail of PM-QKD} \label{Sc:ExpDetail}

\subsection{Laser injection details}
\begin{figure}[htbp]
\includegraphics[width=16cm]{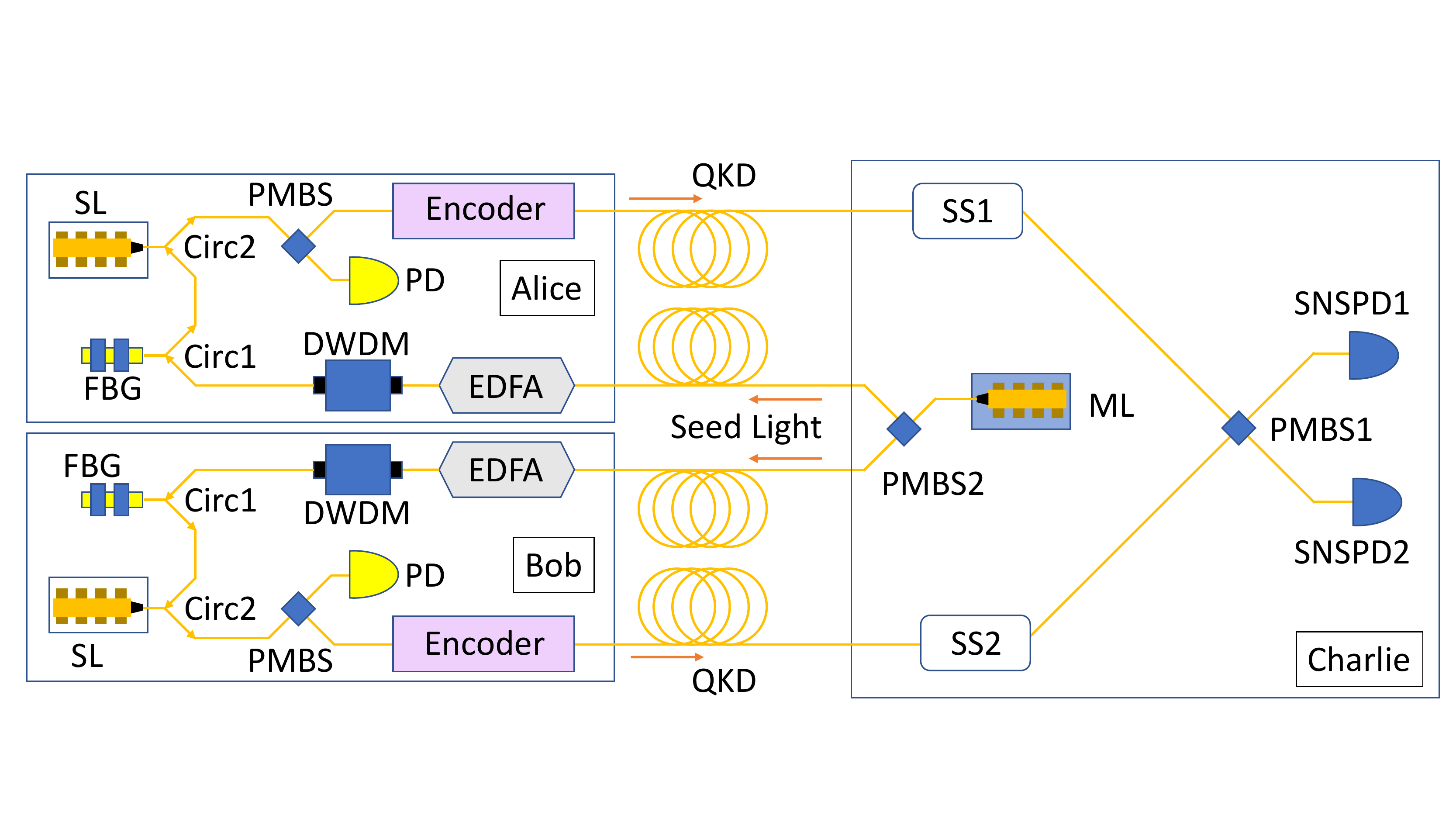}
\caption{Detailed setting of PM-QKD with the nonlocal laser locking system.} \label{fig:nonlocal}
\end{figure}

As is shown in Fig.~\ref{fig:nonlocal}, the continuous wave emitted from the master laser (Realphoton Technology Ltd.) are split by a PMBS and transmitted through long fibres and finally injected into Alice and Bob’s slave laser (Agilecom Ltd.). To achieve a good laser injection result, we apply the Erbium doped fiber amplifier (EDFA) to amplify the light intensity and the dense wavelength division multiplexer (DWDM) to filter the side band noise. After that, a fiber Bragg grating (FBG) inserted before injection is used to block the light with unexpected wavelength. A photoelectric diode (PD) is used to monitor the intensity of injection wave for avoiding that Charlie (or Eve) controls the output of the slave laser by manipulate the injection wave. The optical pulses at the width of 400~ps generated by the slave laser.

\subsection{Encoding and measurement devices}

In accordance with random bit values generated beforehand, two Sagnac rings~\cite{bergh1993simplified} (SR1-2) are employed to modulate the intensities of reference state and three decoy states while two phase modulators (PM1-2) are employed to modulate encoding phases and random phases. The circulator followed are utilized to prevent Trojan horse attack.  Note that the pulse into the SR is split into two pulses by the polarization-maintaining beam splitter (PMBS), one of which travels clockwise and the other travels anticlockwise. The PM in the SR adjusts and loads the phase only on one of the two pulses every round so that the output pulse’s intensity could be modulated when they meet again on the PMBS. To avoid the disturbance on the encoding process, the backward output pulses will be removed by the Circ (circulator) in the SR. Because of the same fiber in which two pulses travel, the clockwise and counter-clockwise paths are insensitive to any length drifts or stress and temperature variation of the fiber within propagation time. In that case, the SR is so stable that the intensity feedback device is unnecessary.

In Charlie's measurement device, to make sure that the polarization of two pulses are indistinguishable, we utilize two PBSs before interfere to filter unexpected part. The reflection port of each PBS is monitored by the SNSPD with detection efficiency of $20\%$ and dark count rate of 100~cps. Charlie adjusts the DC loaded in the EPC in real time to minimize and stabilize the detection count of SNSPD so the incident pulses intensity for interfere maintains maximum. Note that since we have to send strong reference pulses which is described in Appendix \ref{Sc:PhaseEst}, we always have enough pulses intensity for the polarization feedback system even the fiber length is up to 500~km, and Our feedback system works well with less than $1\%$ loss of the incident pulse intensity.

The detector efficiency of SNSPD1 and SNSPD2 is about $40\%$ in 101, 201, 302 and 402~km experiments, while in 502~km experiment, we enhance the efficiency of two detectors to $46\%$ and $75\%$ for better performance. To get a higher interference visibility, we set a $25\%$ non-overlap between the signal and detection window. The insertion loss of Charlie's measurement device is about 1.2 dB.

\subsection{Synchronization systems}

The entire system are synchronized by two arbitrary-function generators (Tektronix, AFG3253) which is not shown in Fig.~1. Three 100~KHz electric signals in phase are modulated by two arbitrary-function generators. The delay between any two of them is adjustable. Alice, Bob and Charlie use one of them to regenerate their own 312.5~MHz system clock, respectively.
To overlap the independent signal pulses from Alice and Bob precisely, we need to develop the time calibration system. First, Charlie measures the arriving time of the optical pulses from Alice and Bob with a time-to-digital converter (TDC), respectively. Then she adjusts the time delay between the electric signals from the arbitrary-function generator according to the calculation of the averaged arriving time. Due to the long optical fiber length drifting, we have to proceed a time calibration every 30 minutes. The total timing calibration precision is about 20~ps which is precise enough for the implement.

\section{Finite-size parameter estimation} \label{Sc:Finite}
Here we list the formulas to estimate the key length $K$, the phase error number, and the corresponding failure probability. The detailed finite-size security analysis of PM-QKD is in Ref.~\cite{Zeng2019PM}.

\subsection{Key length formula}
In the experimental PM-QKD protocol introduced in Appendix \ref{Sc:protocol}, Alice and Bob group the signals by the phase difference $j_s$. For the signals in group $j_s$, the key length is
\begin{equation}
K^{(j_s)} = M^{(j_s)}_\mu [ 1- H(q_\mu^{even(j_s)}) ] - l_{cor}^{(j_s)},
\end{equation}
where $M^{\mu(j_s)}$ is the clicked rounds when Alice and Bob both send out signal lights with $\mu_{i_a} = \mu_{i_b} = \mu/2$ in the group $j_s$. $q_\mu^{even(j_s)}$ is the estimated fraction caused by the even state
\begin{equation}
q_\mu^{even(j_s)} = \frac{M^{even(j_s)}_\mu}{M^{(j_s)}_\mu},
\end{equation}
where $M^{even(j_s)}_\mu$ is clicked rounds caused by even photon component in the $M^{(j_s)}_\mu$ with estimation method introduced in Appendix \ref{Sc:decoy}.

The error correction cost, given by $l_{cor}^{(j_s)}$, can usually by estimated by a function of bit error rate $E^{(j_s)}_\mu$ among the signals of group $j_s$,
\begin{equation}
l_{cor}^{(j_s)} = f M^{(j_s)}_\mu H(E^{(j_s)}_\mu),
\end{equation}
where $f$ is the error correction efficiency, determined by the error rate $E^{(j_s)}_\mu$ and the specific error correction method. If $E^{(j_s)}_\mu$ is too large, the cost $l_{cor}^{(j_s)}$ may be larger than the generated key length $M^{(j_s)}_\mu (1 - H(q_\mu^{even(j_s)}))$. In this case, Alice and Bob directly abandon this group of data.

The overall key length for all the phase group $j_s$ is
\begin{equation}
K = \sum_{j_s = 0}^{D/2-1} \max_{j_s} \{ K^{(j_s)}, 0 \}.
\end{equation}

Note that, the information leakage is only related to the fraction $q_\mu^{even(j_s)}$, irrelevant to the bit error rate. Therefore, the overall information leakage depends on the even photon component in the clicked signals for all the phase group.

In practice, the data post-processing procedure can be improved. Alice and Bob can first perform error correction separately, with respect to different phase group $j_s$. If the error rate is too large, they can simply discard that group of data. Denote the group set $J$ to be the set of group indeces $\{j_s\}$ where the data is left. That is, if $j_s\in J$, then the phase group $j_s$ is kept for key generation. For all the left phase groups in set $J$, Alice and Bob estimate the even photon fraction $q_\mu^{even(J)}$. They then perform privacy amplification among the left data altogether. Of course, Alice and Bob should properly set the group set $J$ to maximize the key length.

With the improved post-processing procedure, the key length for the left phase groups is
\begin{equation}
K = M_\mu^{(J)} [ 1 - H(q^{even(J)}_\mu)] - f \sum_{j_s\in J} M_\mu^{(j_s)} H(E^{(j_s)}_\mu),
\end{equation}
where $M_\mu^{(J)}$, $q^{even(J)}_\mu$ are, respectively, the clicked rounds and estimated even photon fractions for the left data.

\subsection{Decoy-state estimation} \label{Sc:decoy}
The core of parameter estimation step is to estimate the phase error number $M^{even(J)}_\mu$ in the signal state group set $J$ with $\mu_a=\mu_b=\mu/2$. According to the security analysis in Ref.~\cite{Zeng2019PM}, $M^{even(J)}_\mu$ is bounded by
\begin{equation} \label{eq:MXmu}
M^{even(J)}_\mu = 1 - \sum_{k:odd} M_k^{s(J)} \geq 1 - M_1^{s(J)},
\end{equation}
where $M_k^{s(j_s)}$ is the $k$-photon clicked number in the group set $J$ of signal lights.

In experimental PM-QKD protocol, Alice and Bob set the signal, weak decoy and vacuum light with preset ratio $r^{s}, r^{w}, r^{vac}$. To estimate $M_1^{s(J)}$, Alice and Bob perform the following decoy-state process,
\begin{enumerate}
\item To record the sending round number when Alice and Bob both send out signal, weak decoy or vacuum pulses, denoted as $\{N^s, N^w, N^{vac}\}$; and the number of according single-clicked rounds (including the rounds with error), denoted as $\{M^s, M^w, M^{vac}\}$.
\item Based on an inversed usage of Chernoff bound introduced in Appendix \ref{Sc:CHbound}, to calculate $\{ \mathbb{E}^U(\bar{M}^a), \mathbb{E}^L(\bar{M}^a) \}_a$, where $a=s, w, vac$ as signal, weak decoy and vacuum pulse, and estimate the failure probability $\epsilon_1$. Calculate the $\{ \mathbb{E}^U(\bar{Q}^a), \mathbb{E}^L(\bar{Q}^a) \}_a$ by
\begin{equation} \label{eq:Qa}
\bar{Q}^{a*} = \frac{\bar{M}^a}{N^{a}}.
\end{equation}

\item Calculate the minimal $Y_1^{*L}$ based on $\{ \mathbb{E}^U(\bar{Q}^{a*}), \mathbb{E}^L(\bar{Q}^{a*}) \}_a$ and the following formula,
\begin{equation} \label{eq:Y1star}
\bar{Y}^*_1 \geq \bar{Y}^{*L}_1 =\frac{\mu}{\mu\nu - \nu^2} \left( \mathbb{E}^L[\bar{Q}^{w*}]e^\nu - \mathbb{E}^U[\bar{Q}^{s*}]e^\mu \frac{\nu^2}{\mu^2} - \frac{\mu^2 - \nu^2}{\mu^2} \mathbb{E}^U[\bar{Q}^{vac*}] \right),
\end{equation}
calculate $\bar{M}^L_k$ by
\begin{equation} \label{eq:Yk}
\bar{M}_k = N^\infty_k \bar{Y}^*_k ,
\end{equation}
where
\begin{equation}
N^\infty_k = N \sum_{a} P^a(k) (r^a)^2,
\end{equation}
are the asymptotic rounds of sending pulses with photon number $k$. $P^a(k)$ is the Poisson distribution when the intensity setting is $a$.

\item Based on a direct usage of Chernoff bound introduced in Appendix \ref{Sc:CHbound}, to calculate $(M^{s(J)}_1)^L$ and estimate the failure probability $\epsilon_2$.
\end{enumerate}

As a result, Alice and Bob can obtain an lower bound estimation of $M^{s(J)}_1$ with failure probability $\epsilon_1 + \epsilon_2$. From Eq.~\eqref{eq:MXmu}, we can bound the phase error number $M^{even(J)}_\mu$.

\subsection{Chernoff-Hoeffding bound} \label{Sc:CHbound}

Here we present the methods to evaluate $\mathbb{E}(\bar{M}^a)$ from $M^a$ and evaluate $M^s_1$ from $M_1$ using Chernoff bounds.

To evaluate $\mathbb{E}(\bar{M}^a)$ from $M^a$, we inversely use the Chernoff bounds based on Bernoulli variables. We briefly summarize the results in Ref.~\cite{zhang2017improved}. For the observed value $\chi$, we set the lower and upper bound of estimated $\mathbb{E}(\chi)$ as $\{\mathbb{E}^L(\chi), \mathbb{E}^U(\chi)\}$. Denote
\begin{equation}
\begin{aligned}
\mathbb{E}^L(\chi) &= \frac{\chi}{1 + \delta^L}, \\
\mathbb{E}^U(\chi) &= \frac{\chi}{1 - \delta^U}. \\
\end{aligned}
\end{equation}
The failure probability of the estimation $\mathbb{E}(\chi) \in [\mathbb{E}^L(\chi), \mathbb{E}^U(\chi)]$, given by the Chernoff bound, is
\begin{equation} \label{eq:eps1}
\epsilon = e^{-\chi g_2(\delta^L)} + e^{-\chi g_2(\delta^U)},
\end{equation}
where $g_2(x) = \ln(1+x) - x/(1+x)$.

To evaluate $M^s_1$ from $M_1$, we directly apply the Chernoff bounds. Suppose the direct sampling expectation value of $M^s_1$ is given by $\mathbb{E}(M^s_1) = p_1^s M_1$. For the expected value $\mathbb{E}(\chi)$, we set the lower and upper bound of the estimated $\chi$ as $\{\chi^L,\chi^U\}$. Denote
\begin{equation}
\begin{aligned}
\chi^L &= (1 - \bar{\delta}^L)\mathbb{E}(\chi), \\
\chi^U &= (1 + \bar{\delta}^U)\mathbb{E}(\chi). \\
\end{aligned}
\end{equation}
The failure probability of the estimation $\chi \in [\chi^L, \chi^U]$, given by the Chernoff bound, is
\begin{equation} \label{eq:eps2}
\epsilon = e^{-(\bar{\delta}^L)^2\mathbb{E}(\chi)/(2+\bar{\delta}^L)} + e^{-(\bar{\delta}^U)^2\mathbb{E}(\chi)/(2+\bar{\delta}^U)}.
\end{equation}

In practice, we can preset the lower bound and upper bound $\{\mathbb{E}^L(\chi), \mathbb{E}^U(\chi)\}$ or $\{\chi^L,\chi^U\}$ by assuming a Gaussian distribution on $\chi$ first,
\begin{equation}
\begin{aligned}
\mathbb{E}^L(\chi) = \chi - n_\alpha \sqrt{\chi},&\quad \mathbb{E}^U(\chi) = \chi + n_\alpha \sqrt{\chi}, \\
\chi^L = \chi - n_\alpha \sqrt{\mathbb{E}(\chi)},&\quad \chi^U = \chi + n_\alpha \sqrt{\mathbb{E}(\chi)}, \\
\end{aligned}
\end{equation}
where $n_\alpha$ is a preset parameter to determine the estimation precision. After that, we calculate the failure probabilities by Eq.~\eqref{eq:eps1} and Eq.~\eqref{eq:eps2}.

\section{Simulation formulas and detailed results} \label{Sc:simulation}

Here we list the formulas used to simulate the key rate of PM-QKD and MDI-QKD in Fig.~3 in Main Text. The channel is modeled to be a pure loss one and symmetric for Alice and Bob with transmittance $\eta$ (wit the detector efficiency $\eta_d$) taken into account.

\subsection{Gain, yield and error rate of PM-QKD}

In PM-QKD, the $k$-photon the yield $Y_k$ is ( Eq.~(B13) in Ref.~\cite{ma2018phase})
\begin{equation}
Y_k \approx 1 - (1 - 2 p_d) (1-\eta)^k,
\end{equation}
and the gain $Q_{\mu}$ when $\mu_a = \mu_b = \mu/2$, is ( Eq.~(B14) in Ref.~\cite{ma2018phase})
\begin{equation}
Q_\mu \approx 1 - (1 - 2 p_d) e^{-\eta \mu}.
\end{equation}

The even signal fraction $q^{even}_\mu$ is bounded by
\begin{equation}
q^{even}_\mu \leq 1 - Y_1.
\end{equation}

For the coherent lights with intensities $\mu_a - \mu_b = \mu/2$ and phase difference $\phi_a - \phi_b = \phi_\delta$, the bit error rate is ( Eq.~(B21) in Ref.~\cite{ma2018phase} )
\begin{equation} \label{eq:EmuZ}
E_\mu^Z(\phi_\delta) \approx \frac{e^{-\eta\mu}}{Q_\mu} [p_d +\eta\mu \sin^2(\frac{\phi_\delta}{2})].
\end{equation}

In the simulation in Fig.~3 in Main Text, we directly set $e_d(0)=5.3\%$ and only consider the data with matched phases. The bit error rate is,
\begin{equation}
E^Z_\mu(0) \approx \left[ p_d + \eta\mu e_d(0) \right]\frac{e^{-\eta\mu}}{Q_\mu}.
\end{equation}

\subsection{Simulation formulas for MDI-QKD protocols} \label{Sc:simuMDI}

The key rate of MDI-QKD is \cite{Lo2012Measurement}
\begin{equation}
\begin{aligned}
R_{MDI} = \dfrac{1}{2}\{ Q_{11} [1 - H(e_{11})] -f Q_{rect} H(E_{rect}) \},
\end{aligned}
\end{equation}
where $Q_{11} = \mu_a\mu_b e^{-\mu_a-\mu_b}Y_{11}$ and $1/2$ is the basis sifting factor. We take this formula from Eq.~(B27) in Ref.~\cite{Ma2012Alternative}. In simulation, the gain and error rates are given by

\begin{widetext}
\begin{equation}
\begin{aligned}
Y_{11} & = (1-p_d)^2 [ \dfrac{\eta_a\eta_b}{2} + (2\eta_a + 2\eta_b -3\eta_a\eta_b)p_d + 4(1-\eta_a)(1-\eta_b)p_d^2 ], \\
e_{11} & = e_0Y_{11} - (e_0- e_d)(1-p_d^2)\dfrac{\eta_a\eta_b}{2}, \\
Q_{rect} & = Q_{rect}^{(C)} + Q_{rect}^{(E)}, \\
Q_{rect}^{(C)} & = 2(1-p_d)^2 e^{-\mu^\prime/2} [1 - (1-p_d)e^{-\eta_a\mu_a/2}][1 - (1-p_d)e^{-\eta_b\mu_b/2}], \\
Q_{rect}^{(E)} & = 2p_d(1-p_d)^2 e^{-\mu^\prime/2}[I_0(2x) - (1-p_d)e^{-\mu^\prime/2}]; \\
E_{rect} Q_{rect} & = e_d Q_{rect}^{(C)} + (1 - e_d) Q_{rect}^{(E)}, \\
\end{aligned}
\end{equation}
\end{widetext}
Here,
\begin{equation} \label{eqn:MDI mu x}
\begin{aligned}
\mu^\prime & = \eta_a \mu_a + \eta_b \mu_b, \\
x & = \frac12\sqrt{\eta_a \mu_a \eta_b \mu_b}, \\
\end{aligned}
\end{equation}
where $\mu^\prime$ denotes the average number of photons reaching Eve's beam splitter, and  $\mu_a = \mu_b = \mu/2, \eta_a = \eta_b = \eta$. We take these formulas from Eqs.~(A9),~(A11),~(B7),~and (B28)-(B31) in Ref.~\cite{Ma2012Alternative}.

In 2014, Takeoka\textit{~et~al.} derived an upper bound of the key rate of the point-to-point-type QKD protocols \cite{takeoka2014fundamental},
\begin{equation}
R_{TGW} = -\log_2(\dfrac{1-\eta}{1+\eta}).
\end{equation}
Later, Pirandola\textit{~et~al.} established a tight upper bound \cite{Pirandola2017Fundamental},
\begin{equation}\label{eq:plob}
R_{PLOB} = - \log_2(1-\eta),
\end{equation}
which is the linear key-rate bound used in Main Text. 



\section{Experimental data} \label{Sc:ExpData}

Here we list the experiment data for reference. Table~\ref{tab:parameter} and \ref{tab:key} illustrate the data for theoretical parameters, channel condition and the key length calculation. Table~\ref{tab:100-500} show the sending and received statistics of all the signals. Table~\ref{tab:detaileddata1} and Table~\ref{tab:detaileddata2} shows the detailed data of $101$ km experiment with different phase slices setting.

\begin{table}[htbp]
  \centering
  \caption{Parameter setting} \label{tab:parameter}
    \begin{tabular}{lc}
    \hline
    Error correction efficiency & 1.1 \\
    Failure probability & $\sim 1.7\times 10^{-10}$ \\
    Fluctuation factor & 7 \\
    \hline
    \end{tabular}%
\end{table}%

\begin{table}[htbp]
  \centering
  \caption{Channel condition and key generation} \label{tab:key}
    \begin{tabular}{c|ccccc}
    \hline
    \hline
    Distance (km) & 101 & 201 & 302 & 402 & 502  \\
    \hline
    Channel loss   & 1.02E-1 & 1.05E-2 & 1.29E-3 & 1.91E-4 & 8.18E-5 \\
    detection loss   & $23\%$ & $23\%$ & $23\%$ & $20\%$ & $29\%$ \\
    Total loss(double side)   & 2.40E-03 &	2.53E-05 &	3.77E-07 &	7.34E-09 &	1.96E-09 \\
    Dark count   &  2.29E-07 &	5.85E-08 &	7.75E-08 &	3.36E-08 &	1.26E-08 \\
    PLOB bound   &	3.47E-03   &	3.66E-05   &	5.44E-07   &	1.06E-08   &	2.82E-09\\
    \hline
    intensity of decoy state(single side)   &	0.0179   & 	0.0182   & 	0.0192   & 	0.0177   & 	0.0127 	\\
    intensity of signal state(single side)   &	0.0358   & 	0.0364   & 	0.0384   & 	0.0353   & 	0.0253 	\\

    Sending rounds   &	1010250633000   & 	1001472066000   & 	20000133132000   & 	19996635312500   & 	20003396875000 \\
    Aligned bit error rate   &	$5.31\%$   &	$5.75\%$   &	$6.06\%$   &	$7.00\%$   &	$9.80\%$ \\
    Key length   &	98957100   &	6502240   &	13479300   &	287710 & 33674	\\
    Aligned key length   &	48529900   &	3440190   &	7809030   &	287710   &	33674\\
    Expandsion factor   &	2.04   & 	1.89    &	1.73   & 	1.00   & 	1.00 \\
    Failure probablity   &	1.68E-10   &	1.67E-10   &	1.68E-10   &	1.69E-10   &	1.71E-10\\
    Key Rate(bps)    & 2.06E+04   &	1.36E+03   &	9.44E+01   &	2.01E+00   &	1.18E-01\\

    \hline
    \hline

    \end{tabular}%
\end{table}%

\begin{table}[htbp]
  \centering
  \caption{Experimental data of all signals} \label{tab:100-500}
    \begin{tabular}{cc|ccccc}
    \hline
    \hline
    \multicolumn{2}{c|}{Distance (km)} & 101 & 201 & 302 & 402 & 502  \\
    \hline
    \multicolumn{1}{c|}{\multirow{3}{3.5cm}{Sending rounds}} &	Vacuum state &	48107173 & 	453046887 & 25000166415 &	147832268203 &	414356078125 \\
    \multicolumn{ 1}{c|}{} &	Decoy state &	6975540085 &	23129235810 &	280716154317 &	1352629545781 &	3020512928125 \\
    \multicolumn{ 1}{c|}{} & Signal state &	836776167162 & 	684363089673 &	14331523970016 & 	8554989086016 & 	4369313403125 \\

    \hline
    \multicolumn{1}{c|}{\multirow{3}{3.5cm}{Received rounds  (without alighment)}} &	Vacuum state &	22 &	53 &	3875 &	9931 &	10413 \\
    \multicolumn{ 1}{c|}{} &	Decoy state &	6036008 &	2052221 &	3177133 &	1930178 &	1912753\\
    \multicolumn{ 1}{c|}{} & Signal state &	   1365570236 &	120111317 &	325475042 &	23839645 &	5367776\\

    \hline
    \multicolumn{1}{c|}{\multirow{3}{3.5cm}{Received rounds of $j_d = 0$}} &	Vacuum state &	2 &	10 &	430 &	1273 &	1378\\
    \multicolumn{ 1}{c|}{} &	Decoy state &	764808 &	254969 &	397127 &	240748 &	238877\\
    \multicolumn{ 1}{c|}{} & Signal state &	    170644117 &	15018534 &	40698151 &	2982369 &	669910\\

    \hline
    \multicolumn{1}{c|}{\multirow{3}{3.5cm}{Erroneous rounds of $j_d = 0$}} &	Vacuum state & 1 &	5 &	215 &	623 &	689\\
    \multicolumn{ 1}{c|}{} &	Decoy state & 45972 &	15921 &	28982 &	18418 &	35874\\
    \multicolumn{ 1}{c|}{} & Signal state &	9063823 &	863533 &	2467769 &	208910 &	65623\\

    \hline
    \multicolumn{1}{c|}{\multirow{3}{3.5cm}{Received rounds of $j_d = 1$}} &	Vacuum state & 2 &	10 &	502 &	1177 &	1240\\
    \multicolumn{ 1}{c|}{} &	Decoy state & 754770 &	257611 &	398227 &	241664 &	239632\\
    \multicolumn{ 1}{c|}{} & Signal state &	170615800 &	15014075 &	40701711 &	2981797 &	671469\\

    \hline
    \multicolumn{1}{c|}{\multirow{3}{3.5cm}{Erroneous rounds of $j_d = 1$}} &	Vacuum state & 1 &	5 &	251 &	583 &	620\\
    \multicolumn{ 1}{c|}{} &	Decoy state & 62393 &	17727 &	34677 &	28346 &	34094\\
    \multicolumn{ 1}{c|}{} & Signal state &	12678651 &	1098100 &	2997090 &	287194 &	105489\\
    \hline
    \hline
    \end{tabular}%
\end{table}%

\onecolumngrid
\begin{center}
\begin{sidewaystable}
\centering
\caption{Detailed data I of 101 km}
\label{tab:detaileddata1}%
\vspace{0.25cm}
\begin{tabular}{c|cc|cc|cc|cc|cc|cc|cc|cc}
\hline
\hline
\multicolumn{1}{c|}{\multirow{2}{2cm}{Phase Slice}} & \multicolumn{2}{c|}{OPD 0} & \multicolumn{2}{c|}{OPD 1} & \multicolumn{2}{c|}{OPD 2} & \multicolumn{2}{c|}{OPD 3} & \multicolumn{2}{c|}{OPD 4} & \multicolumn{2}{c|}{OPD 5} & \multicolumn{2}{c|}{OPD 6} & \multicolumn{2}{c}{OPD 7} \\
\cline{2-17}
& DETL  & DETR & DETL  & DETR & DETL  & DETR & DETL  & DETR & DETL  & DETR & DETL  & DETR & DETL  & DETR & DETL  & DETR \\
\hline
0 &	375744  &	3969399  &	185800  &	4164734  &	271002  &	4063293  &	615330  &	3766020  &	1196397  &	3213057  &	1844228  &	2403935  &	2539314  &	1633970  &	3081410  &	933133\\
1 &	537284 &	6098291 &	322140 &	6338669 &	513007 &	6265998 &	1125056 &	5733702 &	1979511 &	4666717 &	3035333 &	3521159 &	3986194 &	2309012 &	4963645 &	1354001\\
2 &	435376 &	4551069 &	232993 &	4891727 &	365819 &	4882103 &	791599 &	4348193 &	1468795 &	3648462 &	2223757 &	2723039 &	3093557 &	1878854 &	3745004 &	1062031\\
3 &	414272 &	4708067 &	235091 &	5042261 &	381802 &	4818464 &	839365 &	4361518 &	1484969 &	3562085 &	2337189 &	2748193 &	3107710 &	1806382 &	3874772 &	1042503\\
4 &	430329 &	5080877 &	259194 &	5201990 &	443091 &	5065594 &	911594 &	4466863 &	1673594 &	3780426 &	2497162 &	2796171 &	3416097 &	1894911 &	4085431 &	1034986\\
5 &	422237 &	4868334 &	276350 &	5083603 &	451102 &	4817269 &	968904 &	4409499 &	1672502 &	3595558 &	2565630 &	2742120 &	3372691 &	1768244 &	4170166 &	1027539\\
6 &	421145 &	4748597 &	275524 &	4828410 &	479551 &	4753741 &	958365 &4199669 &	1703840 &	3526144 &	2521076 &	2564003 &	3413853 &	1752309 &	4019548 &	929900\\
7 &	469004 &	5140467 &	320768 &	5437178 &	511651 &	5188944 &	1066672 &	4744950 &	1848169 &	3811145 &	2825253 &	2954097 &	3681213 &	1868314 &	4541764 &	1058875\\
8 &	675823 &	5205534 &	362861 &	5409495 &	497550 &	5391926 &	1025723 &	4774534 &	1906743 &	4130350 &	2908171 &	3028818 &	4056797 &	2089810 &	4947662 &	1177153\\
9 &	751984 &	6139958 &	434122 &	6569900 &	556244 &	6260261 &	1136504 &	5862130 &	2024360 &	4742649 &	3189714 &	3687752 &	4316086 &	2449242 &	5343228 &	1436882\\
10 &	643626 &	4677926 &	298356 &	4877353 &	318169 &	4994816 &	666614 &	4464679 &	1323498 &	3880965 &	2147292 &	2959587 &	3041639 &	2066012 &	3858366 &	1214171\\
11 &	621568 &	4660951 &	301434 &	5115293 &	304863 &	4937140 &	673244 &	4647285 &	1326509 &	3885424 &	2167386 &	3036619 &	3074137 &	2076242 &	3878727 &	1234517\\
12 &	613155 &	4773223 &	270746 &	4953358 &	299752 &	5010456 &	667994 &	4535278 &	1311570 &	3871742 &	2144461 &	2969043 &	3014039 &	2044082 &	3890471 &	1227363\\
13 &	503642 &	4341689 &	236014 &	4707312 &	283689 &	4579890 &	634603 &	4215814 &	1252688 &	3537445 &	2010096 &	2713404 &	2875329 &	1890374 &	3711009 &	1116104\\
14 &	494467 &	4453118 &	239784 &	4642063 &	306664 &	4583053 &	683584 &	4154620 &	1300011 &	3478188 &	2112338 &	2713138 &	3011225 &	1869033 &	3679400 &	1075603\\
15 &	570313 &	5111750 &	287062 &	5418330 &	373937 &	5281927 &	817979 &	4793549 &	1578897 &	4101838 &	2573643 &	3187046 &	3485923 &	2131474 &	4361228 &	1252783\\
\hline
\hline
\end{tabular}
\end{sidewaystable}
\end{center}

\onecolumngrid
\begin{center}
\begin{sidewaystable}
\centering
\caption{Detailed data II of 101 km}
\label{tab:detaileddata2}%
\vspace{0.25cm}
\begin{tabular}{c|cc|cc|cc|cc|cc|cc|cc|cc}
\hline
\hline
\multicolumn{1}{c|}{\multirow{2}{2cm}{Phase Slice}} & \multicolumn{2}{c|}{OPD 8} & \multicolumn{2}{c|}{OPD 9} & \multicolumn{2}{c|}{OPD 10} & \multicolumn{2}{c|}{OPD 11} & \multicolumn{2}{c|}{OPD 12} & \multicolumn{2}{c|}{OPD 13} & \multicolumn{2}{c|}{OPD 14} & \multicolumn{2}{c}{OPD 15} \\
\cline{2-17}
& DETL  & DETR & DETL  & DETR & DETL  & DETR & DETL  & DETR & DETL  & DETR & DETL  & DETR & DETL  & DETR & DETL  & DETR \\
\hline
0 &	3576331 &	462675 &	3698372 &	227367 &	3688156 &	309506 &	3282896 &	668922 &	2818592 &	1292424 &	2062095 &	2012194 &	1419362 &	2827226 &	792026 &	3473682\\
1 &	5472601 &	638783 &	5808385 &	372595 &	5511681 &	554043 &	5077090 &	1193416 &	4055062 &	2127494 &	3099272 &	3322797 &	2013053 &	4443404 &	1148543 &	5447888\\
2 &	4311221 &	508294 &	4439892 &	248592 &	4426118 &	355970 &	3870006 &	784039 &	3288139 &	1504538 &	2438585 &	2351494 &	1647443 &	3244658 &	925502 &	4016093\\
3 &	4281037 &	469627 &	4571805 &	247678 &	4291668 &	359623 &	3937971 &	823162 &	3203771 &	1539792 &	2421228 &	2412460 &	1588343 &	3304423 &	886645 &	4069628\\
4 &	4676434 &	499303 &	4710458 &	249121 &	4648901 &	397609 &	4089283 &	893256 &	3391403 &	1660164 &	2506369 &	2583865 &	1638318 &	3507512 &	924505 &	4404040\\
5 &	4526437 &	447099 &	4780554 &	244226 &	4534353 &	403877 &	4067129 &	894940 &	3314316 &	1652777 &	2439324 &	2547910 &	1621141 &	3530837 &	904895 &	4423635\\
6 &	4560390 &	432379 &	4646425 &	241914 &	4481594 &	403693 &	3960067 &	897943 &	3209832 &	1635793 &	2415660 &	2565777 &	1591514 &	3547321 &	867169 &	4225159\\
7 &	4984117 &	477533 &	5172315 &	272093 &	4938549 &	452297 &	4357280 &	995027 &	3625455 &	1847187 &	2724208 &	2901542 &	1749707 &	3833764 &	983822 &	4681210\\
8 &	5634057 &	564788 &	5895012 &	287022 &	5712734 &	410723 &	5275605 &	912910 &	4492060 &	1761009 &	3354542 &	2699309 &	2285034 &	3710025 &	1318072 &	4493231\\
9 &	6055441 &	695599 &	6325427 &	393846 &	6308054 &	549074 &	5848996 &	1179938 &	4813596 &	2108247 &	3704310 &	3276566 &	2495890 &	4357721 &	1520550 &	5486773\\
10 &	4442634 &	600914 &	4864009 &	287216 &	4957478 &	335079 &	4497613 &	724541 &	3864613 &	1415560 &	2969901 &	2223641 &	2124926 &	3193987 &	1254639 &	3980210\\
11 &	4590026 &	612811 &	5036806 &	289969 &	4908022 &	334996 &	4556779 &	744118 &	3828204 &	1403545 &	3050678 &	2321539 &	2081010 &	3222761 &	1279160 &	4130325\\
12 &	4612559 &	601061 &	4835746 &	288726 &	4818202 &	348755 &	4360038 &	734421 &	3795422 &	1446993 &	2895683 &	2295465 &	2048632 &	3271887 &	1198245 &	4027685\\
13 &	4181277 &	535961 &	4470633 &	271225 &	4335119 &	325771 &	4059576 &	728951 &	3374100 &	1362937 &	2660620 &	2210587 &	1792712 &	3014752 &	1091187 &	3880884\\
14 &	4221721 &	530266 &	4381533 &	271406 &	4394063 &	364214 &	3946303 &	752179 &	3398431 &	1428086 &	2558102 &	2213075 &	1800125 &	3134338 &	1009909 &	3820870\\
15 &	4876569 &	606080 &	5261246 &	332588 &	5087582 &	415528 &	4737267 &	890003 &	3913291 &	1638705 &	3090211 &	2641099 &	2023280 &	3578779 &	1204437 &	4537282\\
\hline
\hline
\end{tabular}
\end{sidewaystable}
\end{center}

\end{appendix}

\bibliography{bibPMExp}

\end{document}